\documentclass[aps,prl,twocolumn,showpacs,superscriptaddress]{revtex4-1}

\usepackage{graphicx}
\usepackage{float}
\usepackage{xcolor}

\begin{document}


\title{Unidirectional ripplopolaron charge transport in a three-terminal microchannel device}

\author{A. O. Badrutdinov}
\email[E-mail: ]{alexbadr@oist.jp}
\affiliation{Okinawa Institute of Science and Technology, Tancha 1919-1, Okinawa 904-0495, Japan}
\author{D. G. Rees}
\email[Current address: ]{Cryogenic Ltd, 6 Acton Park Estate, The Vale, London W3 7QE, United Kingdom}
\affiliation{NCTU-RIKEN Joint Research Laboratory, Institute of Physics, National Chiao Tung University, Hsinchu 300, Taiwan}  
\affiliation{RIKEN CEMS, Wako 351-0198, Japan}
\author{J. Y. Lin}
\affiliation{Okinawa Institute of Science and Technology, Tancha 1919-1, Okinawa 904-0495, Japan}
\author{A. V. Smorodin}
\affiliation{Okinawa Institute of Science and Technology, Tancha 1919-1, Okinawa 904-0495, Japan}
\author{D. Konstantinov}
\affiliation{Okinawa Institute of Science and Technology, Tancha 1919-1, Okinawa 904-0495, Japan}

\date{\today}

\begin{abstract}
We study the transport of surface electrons on superfluid helium through a microchannel structure in which the charge flow splits into two branches, one flowing straight and one turned at 90 degrees. According to Ohm’s law, an equal number of charges should flow into each branch. However, when the electrons are ‘dressed’ by surface excitations (ripplons) to form polaron-like particles with sufficiently large effective mass, all the charge follows the straight path due to momentum conservation. This surface-wave induced transport is analogous to the motion of electrons coupled to surface acoustic waves in semiconductor 2DEGs. 
\end{abstract}


\maketitle

\begin{figure}
\centering
\includegraphics[width=1\columnwidth]{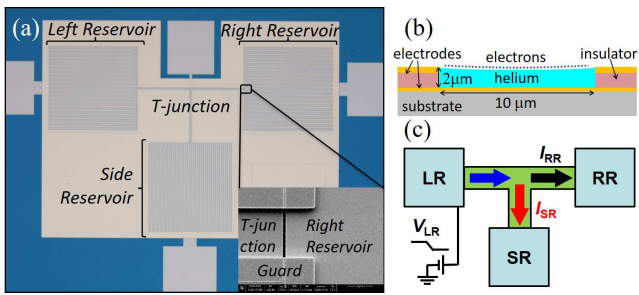}
\caption{\label{fig:1}(color online). Optical microscope image of the microchannel device. Left, right, side reservoirs and the T-junction are indicated. The inset shows a 300 nm gap between the right reservoir and T-junction electrodes. (b) Cross section of the channels forming T-junction. (c) Sketch of the dc measurement technique. The potential of the left reservoir electrode is ramped more negative, as a result, SE currents flow through the T-junction into the right and side reservoirs.} 
\end{figure}

\begin{figure}
\includegraphics[width=1\columnwidth]{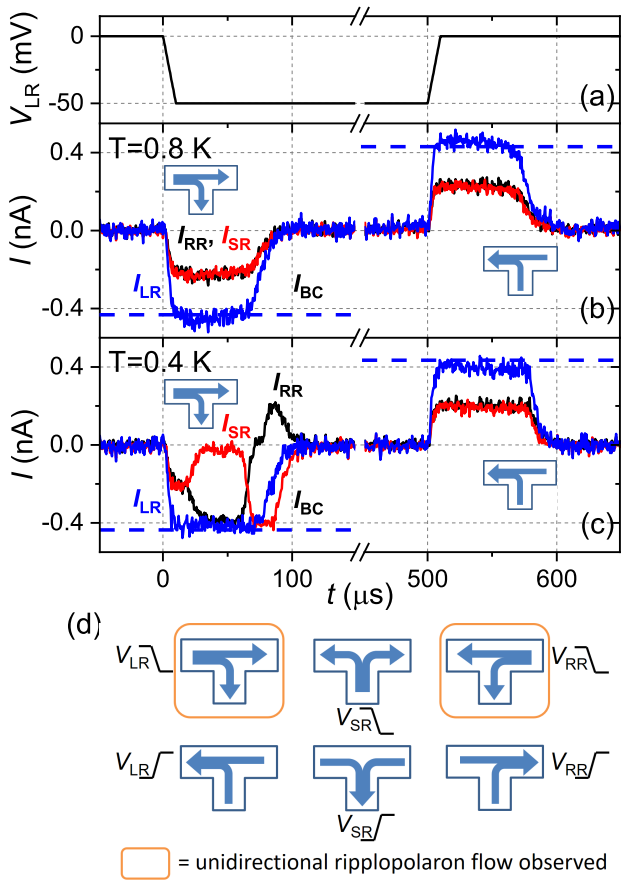}
\caption{\label{fig:2}(color online). Time-resolved SE current measurements taken at density of SE in the T-junction $n_{\rm e}=3.25\times10^{9}$~$\rm cm^{-2}$ and two different values of temperature. (a) Voltage waveform applied to the left reservoir electrode. (b,c) SE currents in the right ($I_{\rm RR}$, black line) and side ($I_{\rm SR}$, red line) reservoirs measured at $T=0.8$ (b) and 0.4~K (c). The current in the left reservoir ($I_{\rm LR}$, blue line) is determined as the sum of $I_{\rm RR}$ and $I_{\rm SR}$, assuming conservation of charge. The dashed lines show calculated values of $I_{\rm BC}=0.43$ $\rm nA$ corresponding to SE current in the BC regime, as described in the text. (d) A diagram, summarizing possible flow configurations for our device, and highlighting configurations where unidirectional ripplopolaron flow is observed at $T=0.4$~K.} 
\end{figure}

Transport properties of charge carriers can be strongly modified by their coupling to medium excitations, for example phonons in solids. The concept of a polaron, that is a self-sustained complex consisting of a localized electron "dressed" by a cloud of medium excitations, was first introduced by Landau and Pekar~\cite{Landau1948} to describe transport of conduction band electrons in ionic crystals. Since then it was extended to many other condensed-matter systems, which include two-dimensional (2D) electron gases in semiconductors~\cite{DasSarma1984,Peeters1986,Peeters1987}, charge carriers in organic materials~\cite{Su1980,Lu2018} and high-Tc superconductors~\cite{Verbist1991,Alex1994}, as well as the Bose-Einstein Condensate (BEC) of cold atoms~\cite{Hu2016,Jorg2016}. Surface electrons (SE) on liquid helium present an exceptionally clean strongly-correlated 2D electron system where a polaronic effect is also important~\cite{Konobook,Andrei}. The polaronic state is formed due to the strong coupling of electrons to the surface medium excitations, the capillary waves (ripplons), and corresponds to an electron self-trapped in a shallow dimple on the surface of the liquid~\cite{Shikin1973,Jackson1981}. This effect is only oberved at high SE densities and low temperatures, for which the electrons form a 2D crystal (Wigner solid)~\cite{Monarkha1975}. In this case, the localisation of electrons at specific lattice sites prevents the thermally induced detrapping from the dimple and consequent destruction of the polaronic state~\cite{Jackson1982,Tress1996,Rubin1999}. 
\newline
\indent In this Letter, we report a novel regime of polaronic charge transport for SE on superfluid helium confined in a three-terminal microchannel device. In this device, the polaron lattice is driven towards a microchannel T-shaped junction where one possible path is straight and the other involves a 90-degree turn. The bare charge current should split equally at the junction following Ohm's law, while for surface waves it is natural to preserve the direction of motion due to momentum conservation. For the polaron lattice, we observe a transport regime in which the influence of momentum conservation dominates and all the charge carriers are carried straight through the junction. This unidirectional ripplopolaron transport is observed when the ripplopolaron mass is enhanced significantly due to the resonant excitation of ripplons by the electron motion, and at temperatures for which the ripplon damping length is sufficiently long. This novel regime of electron transport can find useful applications, in particular for mobile spin qubits on superfluid helium. 
\newline
\indent An optical microscope image of our device is shown in Fig.~\ref{fig:1}(a). The device consists of 2~$\mu$m deep channels fabricated on a silicon dioxide substrate using optical lithography. Three identical sets of 20~$\mu$m wide channels connected in parallel form left, right and side electron reservoirs. Each reservoir has a 10 $\mu$m wide and 400 $\mu$m long microchannel segment attached, and these three segments are connected together forming a three-terminal T-shaped microchannel structure, which we will call the T-junction. The device is placed in a vacuum-tight copper cell cooled down in a dilution refrigerator and partially filled with liquid $^4$He. The device plane is placed slightly above the surface of the liquid in the cell, so the channels are filled with superfluid helium due to capillary action. Electrons are produced by thermionic emission from a heated tungsten filament located above the device and are trapped on the surface of the superfluid helium filling the channels.
The electrostatic potential in each set of channels is controlled by several electrodes organized in two layers (Fig.~\ref{fig:1}(b)). The bottom layer consists of four separate electrodes covering the bottom of the channels in each of three reservoirs and of the T-junction. These electrodes are denoted as \textit{left, right} and \textit{side reservoir} electrodes, and the \textit{T-junction} electrode. The potentials of these electrodes with respect to ground are denoted as $V_{\rm LR}$, $V_{\rm RR}$, $V_{\rm SR}$ and $V_{\rm T}$, respectively. Adjacent electrodes are separated by a 300~nm gap (inset of Fig.~\ref{fig:1}(a)). The top layer constitutes a single electrode, covering the top of the channel grating and denoted as the \textit{guard} electrode, with potential $V_{\rm Gu}$. SE are confined in the channels when the bottom layer is biased more positive than the top layer. The electrons located in the T-junction form the conduction path between the reservoirs. In equilibrium, the SE potential $V_{\rm SE}$ is the same for electrons in all three reservoirs, as well as in the T-junction, while the areal density of SE in each part of the device is determined by potential of the corresponding bottom-layer electrode. The number of charges in the T-junction can be decreased to zero by lowering the potential $V_{\rm T}$ to a certain threshold value, which is determined experimentally from the complete cut-off of SE current through the T-junction. The device therefore resembles a field-effect transistor (FET)~\cite{Klier2000}. The measured threshold value of $V_{\rm T}$ determines the value of $V_{\rm SE}$. The density of SE in the T-junction for an arbitrary value of $V_{\rm T}$ is determined by finite element modeling, as described previously~\cite{Badr2016}. The SE density profile across the channel is calculated for a given SE potential $V_{\rm SE}$. The SE density is then defined as the average value $n_{\rm e}=(1/w)\int n_{\rm e}(y)dy$, where $w$ is the width of the electron system in the channel and $y$ is the direction across the channel. The melting temperature of the WS can then be estimated as $T_m=e^2\sqrt{\pi n_{\rm e}}/(4\pi\varepsilon_0 k_B \Gamma)$, where $e$ is the elementary charge, $\varepsilon_0$ is the permittivity of vacuum, $k_B$ is the Boltzmann constant, and $\Gamma\sim 130$~\cite{Grim1979}. For all measurements reported here, $T_m$ was significantly above 1~K. 
\newline
\indent DC transport of SE in the device was measured using a capacitive-coupling method described in Ref.~\cite{Rees2016}. Initially all reservoirs are set at ground potential, with $V_{\rm Gu}$ more negative and $V_{\rm T}$ sufficiently positive to open the conduction path between the reservoirs. The potential of one of the reservoir electrodes is then linearly ramped to a certain value. As a result, the potential of SE in this reservoir changes and SE flow between this reservoir and the two other reservoirs through the T-junction (Fig.~\ref{fig:1}(c)). The image-charge currents induced in the two other reservoir electrodes by the SE movement are recorded using current preamplifiers and a multi-channel digital storage oscilloscope. The current of SE into each reservoir is calculated from the measured image-charge current by taking into account the capacitive coupling of SE to the reservoir electrode and the guard electrode. The SE current flowing out of the ramped reservoir is determined as the sum of these currents, assuming conservation of charge in the T-junction.
\newline
\indent Results of such measurements made at $T=0.8$~K and at $T=0.4$~K are shown in Fig.~\ref{fig:2}(b) and Fig.~\ref{fig:2}(c), respectively. At $t=0$, the potential of the left reservoir $V_{\rm LR}$ is ramped to $-50$~mV in 10~$\mu$s (Fig.~\ref{fig:2}(a)), causing SE to flow out of the left reservoir, through the T-junction, and into the right and side reservoirs. At $t=500~\mu$s, after all transient currents have stopped and a new equilibrium established, the potential of the left reservoir is ramped back to zero at the same rate (Fig.~\ref{fig:2}(a)), causing SE to flow back into the left reservoir and returning the system to its initial state. In the case of SE flowing out of the left reservoir, there is a striking difference in the current flow observed at $0.8$~K and at $0.4$~K. At $0.8$~K, for the whole duration of flow, $I_{\rm RR}$ and $I_{\rm SR}$ are equal, indicating that at the T-junction SE flow splits equally. In contrast, at $0.4$~K $I_{\rm RR}$ and $I_{\rm SR}$ are initially equal but, after some $20$~$\mu$s, $I_{\rm RR}$ starts to grow and eventually becomes almost equal to $I_{\rm LR}$, while $I_{\rm SR}$ becomes nearly zero. A unidirectional flow through the T-junction is established and lasts for about $40$~$\mu$s. During this period an excess of charge ($\sim1\times10^5e$) builds up in the right reservoir, causing an electrostatic potential difference of some 15 mV to appear between charges in the right and side reservoirs, a strongly non-Ohmic behaviour. After that, a reverse current from the right reservoir into the side reservoir occurs to establish the electrostatic equilibrium with the electrostatic potential in all reservoirs equal.
\newline
\indent This unidirectional flow appears only for specific flow configurations. With our device we could study six distinct flow configurations in total (Fig.~\ref{fig:2}(d)). When the potential of any of the reservoir electrodes is ramped to a more negative value, SE are pushed out of this reservoir towards the junction, and at the junction the SE flow splits. When the potential of any of the reservoir electrodes is ramped less negative, SE are pulled into this reservoir out of two other reservoirs, so that at the junction two SE flows merge into one. At $0.8$~K, the splitting or merging flows are always equal for all possible flow configurations, the expected Ohmic result. At $0.4$~K, the unidirectional flow is observed, but only if SE are flowing out of either the left or right reservoir towards the junction. In the case of SE flowing out of the side reservoir, as well as for all merging flow configurations, the splitting or merging flows are equal.
\newline
\indent An understanding of the interaction between the electron lattice and the helium substrate is crucial in understanding this behaviour. In the case of uniform WS motion, the polaronic effect can be significantly enhanced due to the coherent emission of ripplons with wave vectors equal to the reciprocal lattice vectors $\textbf{G}$. The resonant condition is met whenever the velocity of the WS approaches the phase velocity $v_G=\omega_G/G$ of the emitted ripplons, where $\omega_G=\sqrt{\alpha/\rho}G^{3/2}$ ($\alpha$ and $\rho$ are the surface tension and density of liquid helium, respectively) is the dispersion relation for the capillary waves. This corresponds to a regime of so-called Bragg-Cherenkov (BC) scattering, in which the polaron effective mass is strongly enhanced by the resonant ripplons and the polaron lattice moves with the terminal velocity $v_G$ regardless of the magnitude of the driving force~\cite{Shir1995,Kris1996,Dykm1997,Ikeg2009}. In Figs.~\ref{fig:2}(b,c), shortly after each ramp has started, $I_{\rm LR}$ saturates at a value very close to the value of SE current expected for the BC regime ($I_{\rm BC}$). The latter is calculated as $I_{\rm BC}=v_1en_{\rm e}w$ and represented by the dashed lines in Figs.~\ref{fig:2}(b,c). Here $v_{\rm 1}=\sqrt{\alpha G_1/\rho}$ is the phase velocity of the resonant ripplons with wave vector equal to the first reciprocal lattice vector of the WS, the magnitude of which is related to the SE density as $G_1 = (8\pi^2n_{\rm e}/\sqrt{3})^{1/2}$~\cite{Grim1979}. Because the current is limited to this specific value, the time during which charge can move through the device to establish a new equilibrium is much longer than the duration of the voltage ramp~\cite{Rees2016}. Following this analysis, it is clear that the ripplopolarons travelling from the left reservoir towards the conjunction move with a resonantly enhanced effective mass. The SE in the other two segments move away from the conjunction at a velocity that is twice smaller (when the currents in these two sections are equal). In this case the conditions for resonant BC scattering are not satisfied and the ripplopolaronic effective mass is not enhanced. 
\newline
\indent An intuitive explanation emerges that the unidirectional transport of SE in the T-junction occurs when the BC resonance causes the ripplopolaron mass to become sufficiently large. When SE approach the conjunction of the three segments heavily dressed by the ripplon system, the latter would naturally propagate straight through the conjunction and, being strongly coupled to the SE, can carry the SE with it. When SE flow out of the left or right reservoir, the straight path is available at the conjunction, so the effect can be observed. When SE flow out of the side reservoir there is no straight path available at the conjunction, and the current must split left and right. The unidirectional effect is therefore not observed. In this case we assume that the ripplons dissipate due to collisions with the walls of the T-channel. In the case of merging flow configurations, when SE are pulled into one of the reservoirs out of the two others, the BC resonance occurs for the ripplopolarons as they move away from the conjunction and so does not effect the transport of charge carriers through it. 
\begin{figure}
\includegraphics[width=1\columnwidth]{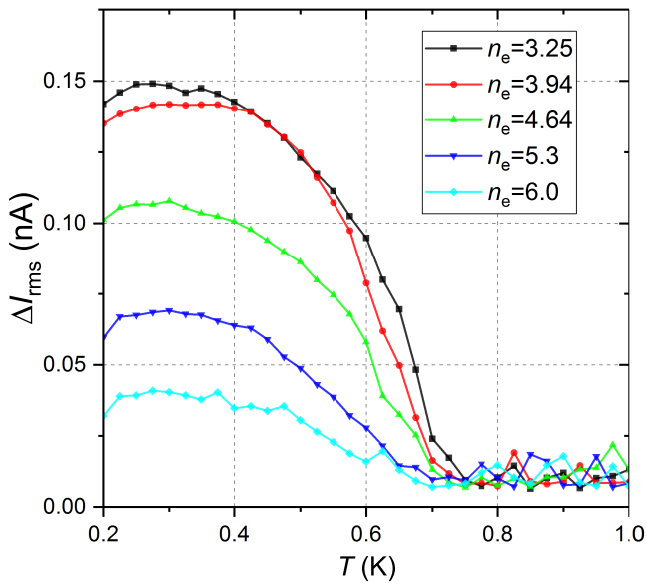}
\caption{\label{fig:3}(color online). $\Delta I_{\rm rms}$ against temperature for different densities of SE in the T-junction. The corresponding densities are indicated in the legend in units of $10^9$~cm$^{-2}$.} 
\end{figure}
\newline
\indent We have carried out an extensive study of this effect by varying a number of experimental parameters, such as the temperature, density of SE, and rate of voltage ramp. We conclude that the crucial parameter for the appearance of the unidirectional transport is temperature: the asymmetry in current through the T-junction appears only below about 0.7~K. Fig.~\ref{fig:3} shows the summary of our data. Each point in this plot is obtained by applying the voltage ramp shown in Fig.~\ref{fig:2}(a) to the left reservoir and recording the current traces similar to the ones shown in Fig.~\ref{fig:2}(b,c). In order to quantify the asymmetry between currents $I_{\rm RR}$ and $I_{\rm SR}$, we introduce the quantity $\Delta I_{\rm rms}$ defined as the rms time-averaged value of $(I_{\rm RR}(t) - I_{\rm SR}(t))$. This quantity is plotted in Fig.~\ref{fig:3} as a function of temperature for several values of the density of SE in the T-junction. The current asymmetry appears below the onset temperature of about 0.7~K, which is only weakly dependent on $n_{\rm e}$. Below the onset temperature, $\Delta I_{\rm rms}$ increases with decreasing temperature and saturates below about 0.4~K. Unlike the value of the onset temperature, the saturation value of $\Delta I_{\rm rms}$ shows a dependence on $n_{\rm e}$. This is because the total number of SE flowing through the junction is determined only by the amplitude of voltage ramp, while the total current increases with density as $n^{5/4}_{\rm e}$. Therefore, the duration of the transient current response decreases with increasing $n_{\rm e}$, which results in shorter duration of the asymmetric flow and lower value of $\Delta I_{\rm rms}$. 
\newline
\indent The observed temperature dependence of the unidirectional flow can be understood as follows. In the BC regime, the resonantly enhanced mass of the ripplopolaron lattice leads to large effective resistance~\cite{Ikeg2009}. As current flows towards, and then splits at, the conjunction, the applied driving potential therefore drops mostly across the T-junction segment adjacent to the driven reservoir, where SE flow in the BC regime. As the ripplopolaron lattice enters the conjunction, the electron velocity and the driving electric field drops and the ripplons, which constitute the lattice, start dissipating due to interaction with bulk excitations in superfluid helium. For unidirectional flow, the crucial quantity is the propagation length of resonant ripplons, which is finite due to their damping and which should be compared to the size of the device. The damping rate of ripplons in superfluid helium was studied by Roach {\it et al.} using an interdigital capacitor device~\cite{Roch1995}. Using the theoretical expressions established therein, the inverse propagation length of resonant ripplons with the wave number $G_1$ can be written as 
\begin{equation}
\lambda^{-1} = \frac{\pi^2}{90}\frac{\hbar G^2_1}{\rho \omega_{G_1}} \left( \frac{k_B T}{\hbar s} \right)^4 \approx 11.7 n^{\frac{1}{4}}_{\rm e} T^4,
\label{eq1}
\end{equation}
where $s$ is the first sound velocity of liquid helium. The propagation length has a weak dependence on SE density but increases rapidly with decreasing temperature, from about 30~$\mu$m at $T=1$~K to about $2\times10^4$~$\mu$m at $T=0.2$~K, for $n_{\rm e} = 4\times 10^9$~cm$^{-2}$. At lower temperatures, when the propagation length of resonant ripplons is comparable to or larger than the length of the segment (400~$\mu$m), the polaronic dimples can carry SE straight through the junction, establishing the unidirectional charge transport. At higher temperatures, the resonant ripplons dissipate over a distance much shorter than the length of the segment, and the diffusive transport of the bare SE obeys Ohm's law. We believe that other temperature-dependent effects, such as vapor-atom scattering, finite concentration of $^3$He atoms on the surface of superfluid helium~\cite{supplemental,And1966,Zin1969,Guo1971,Esel1980}, etc., are irrelevant because, to the best of our knowledge, they do not influence the BC scattering mechanism, ripplopolaron formation or ripplon lifetime.
\newline
\indent Finally, our observation that the unidirectional ripplopolaron flow is established some time after the flow onset suggests that the enhancement of the ripplopolaron mass, that occurs due to BC scattering when the system starts to move, is not immediate. As the SE start to move at the resonant velocity $v_1$ the lattice depth starts growing and, once the effective mass becomes sufficiently high, the unidirectional flow is established~\cite{supplemental}. 
\newline
\indent In summary, we have observed a new transport regime of SE on superfluid helium by employing a specially designed T-shaped microchannel device. The charge carriers in our device are carried straight through the T-junction by resonantly excited ripplons on the surface of superfluid. This transport mechanism is reminiscent of 2D electrons in semiconductors carried by surface acoustic waves (SAW)~\cite{Herm2011,Mac2011}. This suggests a very interesting possibility to use electro-mechanically generated surface waves \cite{Roch1995} on superfluid helium to carry SE without employing driving electric fields. Similar to the flying qubits in semiconductors~\cite{Bert2016,Taka2019}, this technique might be potentially used for mobile spin qubits based on electrons on helium~\cite{Lyon2006,Schu2010,Brad2011,Kool2019,Kawa2019}.  
\newline
\indent The authors are supported by an internal grant from the Okinawa Institute of Science and Technology (OIST) Graduate University. A.O.B. is supported by JSPS KAKENHI Grant No. JP18K13506. D.G.R. was supported by the Taiwan Ministry of Science and Technology (MOST) through Grant No. MOST 103-2112-M-009-017 and the MOE ATU Program.

\bibliography{IEpaperbib}
  
\end{document}